\documentstyle[epsfig]{mn}

\newif\ifAMStwofonts

\def\msun{M$_\odot$}
\def\be7{$^{7}$Be}
\def\n13{$^{13}$N}
\def\o14{$^{14}$O}
\def\oo15{$^{15}$O}
\def\f17{$^{17}$F}
\def\ff18{$^{18}$F}
\def\na22{$^{22}$Na}
\def\al26{$^{26}$Al}

\ifoldfss
  \ifCUPmtlplainloaded \else
    \NewTextAlphabet{textbfit} {cmbxti10} {}
    \NewTextAlphabet{textbfss} {cmssbx10} {}
    \NewMathAlphabet{mathbfit} {cmbxti10} {} 
    \NewMathAlphabet{mathbfss} {cmssbx10} {} 
  \fi
  \ifAMStwofonts
    \ifCUPmtlplainloaded \else
      \NewSymbolFont{upmath} {eurm10}
      \NewSymbolFont{AMSa} {msam10}
      \NewMathSymbol{\upi}     {0}{upmath}{19}
      \NewMathSymbol{\umu}     {0}{upmath}{16}
      \NewMathSymbol{\upartial}{0}{upmath}{40}
      \NewMathSymbol{\leqslant}{3}{AMSa}{36}
      \NewMathSymbol{\geqslant}{3}{AMSa}{3E}

      \let\leq=\leqslant 
       
    \fi
  \fi
\fi 

\ifnfssone
  \newmathalphabet{\mathit}
  \addtoversion{normal}{\mathit}{cmr}{m}{it}
  \addtoversion{bold}{\mathit}{cmr}{bx}{it}
  \newmathalphabet{\mathbfit} 
  \addtoversion{normal}{\mathbfit}{cmr}{bx}{it}
  \addtoversion{bold}{\mathbfit}{cmr}{bx}{it}
  \newmathalphabet{\mathbfss} 
  \addtoversion{normal}{\mathbfss}{cmss}{bx}{n}
  \addtoversion{bold}{\mathbfss}{cmss}{bx}{n}
  \ifAMStwofonts
    \ifCUPmtlplainloaded \else
      \UseAMStwoboldmath
      \makeatletter
      \new@mathgroup\upmath@group
      \define@mathgroup\mv@normal\upmath@group{eur}{m}{n}
      \define@mathgroup\mv@bold\upmath@group{eur}{b}{n}
      \edef\UPM{\hexnumber\upmath@group}
      \new@mathgroup\amsa@group
      \define@mathgroup\mv@normal\amsa@group{msa}{m}{n}
      \define@mathgroup\mv@bold\amsa@group{msa}{m}{n}
      \edef\AMSa{\hexnumber\amsa@group}
      \makeatother
      \mathchardef\upi="0\UPM19
      \mathchardef\umu="0\UPM16
      \mathchardef\upartial="0\UPM40
      \mathchardef\leqslant="3\AMSa36
      \mathchardef\geqslant="3\AMSa3E

      \let\leq=\leqslant 

    \fi
  \fi
\fi 

\ifnfsstwo
  \DeclareMathAlphabet{\mathbfit}{OT1}{cmr}{bx}{it}
  \SetMathAlphabet\mathbfit{bold}{OT1}{cmr}{bx}{it}
  \DeclareMathAlphabet{\mathbfss}{OT1}{cmss}{bx}{n}
  \SetMathAlphabet\mathbfss{bold}{OT1}{cmss}{bx}{n}
  \ifAMStwofonts
    \ifCUPmtlplainloaded \else
      \DeclareSymbolFont{UPM}{U}{eur}{m}{n}
      \SetSymbolFont{UPM}{bold}{U}{eur}{b}{n}
      \DeclareSymbolFont{AMSa}{U}{msa}{m}{n}
      \DeclareMathSymbol{\upi}{0}{UPM}{"19}
      \DeclareMathSymbol{\umu}{0}{UPM}{"16}
      \DeclareMathSymbol{\upartial}{0}{UPM}{"40}
      \DeclareMathSymbol{\leqslant}{3}{AMSa}{"36}
      \DeclareMathSymbol{\geqslant}{3}{AMSa}{"3E}

      \let\leq=\leqslant 

    \fi
  \fi
\fi 

\ifCUPmtlplainloaded \else
  \ifAMStwofonts \else 
    \def\upi{\pi}
    \def\umu{\mu}
    \def\upartial{\partial}
  \fi
\fi

\title[Gamma--ray emission from individual classical novae]
      {Gamma--ray emission from individual classical novae}
\author[G\'omez-Gomar, Hernanz, Jos\'e and Isern] 
        {Jordi G\'omez-Gomar,$^{1}$
        Margarita Hernanz,$^{1}$ 
        Jordi Jos\'e$^{1,2}$ and
        Jordi Isern$^{1}$ \\
$^{1}$ Institut d'Estudis Espacials de Catalunya, 
       CSIC Research Unit, Edifici Nexus-104, 
       C/Gran Capit\`a, 2-4, E-08034 Barcelona, SPAIN \\
$^{2}$ Departament de F\'\i sica i Enginyeria Nuclear (UPC), 
       Avda. V\'\i ctor Balaguer, s/n, 
       E-08800 Vilanova i la Geltr\'u (Barcelona), SPAIN}

\pagerange{\pageref{firstpage}--\pageref{lastpage}}
\pubyear{1997}

\begin{document}

\maketitle

\label{firstpage}

\begin{abstract}
Classical novae are important producers of radioactive nuclei, such as 
\be7, \n13, \ff18, \na22 and \al26. The disintegration of these nuclei 
produces positrons (except for \be7) that through annihilation 
with electrons produce photons of energies 
511 keV and below. Furthermore, \be7 and \na22 decay 
producing  
photons with energies of 478 keV and 1275 keV, respectively, well in the 
$\gamma$-ray domain. Therefore, novae are potential sources of $\gamma$-ray 
emission. 
We have developed two codes in order to analyze carefully the $\gamma$-ray 
emission of individual classical novae: a hydrodynamical one, which follows 
both 
the accretion and the explosion stages, and a Monte-Carlo one, able to treat 
both the production and the transfer of $\gamma$-ray photons. Both codes have 
been coupled in order to simulate realistic explosions. The properties of 
$\gamma$-ray spectra and $\gamma$-ray light curves (for the continuum and for 
the lines at 511, 478 and 1275 keV) have been analyzed, with a special 
emphasis on the difference between carbon-oxygen and oxygen-neon novae. 
Predictions of detectability of individual novae 
by the future SPI spectrometer on board the 
INTEGRAL satellite are made.
Concerning \al26, its decay produces photons of 1809 keV but 
it occurs on a timescale much longer than the typical 
time interval between nova outbursts in the Galaxy, making it undetectable 
in individual novae. The accumulated emission of \al26 from many Galactic
novae has not been modeled in this paper.
\end{abstract}


\section{Introduction}

Thermonuclear runaways on accreting white dwarfs are at the origin of nova
explosions. Some unstable proton-rich nuclei, such as \n13, \o14, \oo15,
\f17 and \ff18, are produced during hydrogen--burning. These nuclei are
$\beta^{+}$-unstable and therefore decay through the emission of positrons. This
has two important consequences. First of all, the decay times of \o14 
($\tau$=102s), \oo15 ($\tau$=176s) and \f17 ($\tau$=93s) 
are such that these nuclei can be transported by convection to the
outer layers of the envelope (during the runaway) before decaying; their
subsequent decay in these outer shells of the envelope is largely
responsible for the expansion and of the large increase in luminosity of the
nova. Second, the positrons emitted by all those nuclei annihilate with
electrons leading to the emission of photons with energy equal or below
511 keV. \n13 ($\tau$=862s) and \ff18 ($\tau$=158min) are the most important 
contributors to $\gamma$-ray emission, since their decay
timescales are larger and they decay when the envelope is transparent enough.
Other radioactive elements are synthesized during nova explosions: 
carbon-oxygen (CO) novae are important producers of \be7 ($\tau$=77days) 
(see Hernanz et al. 1996), which emits a photon of 
478 keV when it decays through an electron capture, whereas 
oxygen-neon-magnesium novae (ONeMg) produce \na22 ($\tau$=3.75yr) and 
\al26 ($\tau$=1.04$\times 10^6$yr), which decay emitting a positron and a 
photon of 1275 and 1809 keV, respectively.

Some previous works have pointed out the potential importance of classical
novae as $\gamma$-ray emitters (Clayton 1981, Clayton \& Hoyle 1974, 
Leising \& Clayton 1987). Concerning the nucleosynthesis of radioactive 
nuclei in novae, it  
has been analyzed by means of parametrized studies (Weiss \& Truran 1990, 
Nofar, Shaviv \& Starrfield 1991), and by complete hydrodynamic simulations 
(see Politano et al. 1995 and Prialnik \& Kovetz 1997, for recent works 
concerning ONeMg and CO novae, respectively, and also Jos\'e \& Hernanz 
1997). But there is a lack of detailed studies of the $\gamma$-ray output 
of novae and its relation with the production of radioactive nuclei. 
Only the use of realistic profiles of velocities, 
densities and chemical abundances, necessary to know the production and 
transfer of $\gamma$-rays during nova explosions, will provide the correct 
$\gamma$-ray spectra of classical novae (Hernanz et al. 1997a and b).

From the observational point of view, two types of observations have to be 
distinguished: individual and cumulative. The emission of short and 
medium-lived radioactive nuclei (478, 511 and 1275 keV lines and continuum) 
can be detected in individual novae, provided they are close enough. 
The emission of very long-lived \al26 nuclei can contribute to the 
diffuse emission in the Galaxy, but can not be detected in particular 
novae. The emission of the medium-lived nuclei (\be7 and \na22 at 478 
and 1275 keV) can also be accumulated from different novae, as their 
lifetimes are longer than the typical time interval between two consecutive 
nova explosions in the Galaxy. 
Up to now, no positive detection of novae in $\gamma$-rays has 
been produced, neither for particular novae in the 478 or 1275 keV lines 
(Harris, Leising \& Share 1991, Iyudin et al. 1995), nor for the 
cumulative emission 
of novae at the Galactic center in these same lines (Leising et al. 1988,  
Harris et al. 1991 and 1996).

We will concentrate in this paper on the $\gamma$-ray 
emission from individual novae, by means of hydrodynamical simulations which 
include 
both the accretion and explosion stages, with complete nucleosynthesis, and a 
Monte-Carlo technique to follow the production and transfer of $\gamma$-ray 
photons. The analysis of the cumulative emission from novae in the Galaxy is 
currently under way.
The nova models are presented in section 2, with special emphasis 
on the synthesis of radioactive nuclei and their density and velocity 
profiles. In section 3 we describe the Monte-Carlo simulation technique and 
in section 4  
its results, in the form of $\gamma$-ray spectra and light curves for the 
continuum and the main lines of interest. Finally, our discussion and 
conclusions follow.

\section{Nucleosynthesis of radioactive nuclei}

The accretion and explosion stages have been followed by means of an implicit 
hydrodynamic code, which includes a complete nuclear reaction network; 
one hundred nuclei, ranging from $^1$H to $^{40}$Ca, linked through 
370 nuclear reactions, have been included (see Jos\'e 1996 and Jos\'e \& 
Hernanz 1997a and b, for details). A time-dependent formalism for convective 
transport has been included whenever the characteristic convective timescale 
becomes larger than the integration time step (Wood 1974). Partial mixing 
between adjacent shells is treated as a diffusion process (Prialnik, Shara 
\& Shaviv 1979). CO as well as ONeMg novae have been computed, 
with an accretion rate of 2$\times 10^{-10}$ \msun yr$^{-1}$ (other values 
for the accretion rate have also been considered, but they do not introduce 
noticeable changes in the properties of the radioactive nuclei synthesized). 
The initial 
composition of nova envelopes is far from being understood, but some 
enrichment with material from the white dwarf core is needed in order to 
explain both the explosion mechanism itself and some of the observed abundances 
in novae. Thus, as the 
real mechanism of enhancement is not clear at present, a compromise solution 
consisting of a parametrized mixture between the 
outermost shells of the underlying core and the solar-like accreted matter
is adopted. We have assumed a 50\% degree of 
mixing, although other possibilities have also been analyzed. For the CO 
novae, the adopted composition of the underlying core consists of a 
50\%-50\% mixture of 
$^{12}$C and of $^{16}$O, whereas for ONeMg 
novae (where the initial composition is more crucial for the synthesis of 
the radioactive nucleus \na22) we have adopted a detailed outer core initial 
composition from Ritossa, Garc\'{\i}a-Berro \& Iben (1996) (indicating 
that these white dwarfs are almost devoid of magnesium, which makes it more 
correct to call them oxygen-neon (ONe) white 
dwarfs). The exact initial composition is an important issue, since the 
presence of $^{20}$Ne and its amount, as well as that of magnesium, is crucial 
for the final amount of the radioactive nuclei \na22 and \al26. In particular, 
the 
new initial abundances we are using favor a lower \na22 and \al26 production 
than the ones commonly adopted, based on old nucleosynthesis calculations by 
Arnett \& Truran (1969) (Jos\'e, Hernanz \& Coc 1997, Jos\'e \& Hernanz 
1997a and b). 
Recent observations of individual novae with the COMPTEL instrument 
on board the COMPTON Gamma-Ray Observatory (CGRO) have not detected 
the 1275 keV emission in any of the observed novae. Only upper limits to the flux, 
which translate into upper limits to the \na22 ejected mass, were 
obtained (Iyudin et al. 1995). COMPTEL observations have also 
provided detailed maps of the Galactic 1809 keV emission, corresponding 
to \al26 decay (Diehl et al. 1995). The main features of these maps 
seem to indicate that this emission is rather correlated with a young
population, thus discarding novae as the main contributors to it 
(see Prantzos \& Diehl 1996 for a recent review). Our low \na22 and 
\al26 yields are consistent with both observational results.

The production of radioactive nuclei, as well as some other relevant 
properties, of some selected CO and ONe novae are displayed in Table 
\ref{Tab1} (see Jos\'e and Hernanz 1997a and b for more details). 
Models A1 and A2 correspond to ONe novae with masses 1.15 \msun~and 
1.25 \msun, 
respectively, whereas models B1 and B2 correspond to CO novae, with masses 
0.8 and 1.15 \msun , respectively. 
Models A1, A2 and B2 share some general properties: they have similar 
ejected masses ($\sim 10^{-5}$ \msun) and similar kinetic energies 
($\sim 3\times 10^{16}$ erg.g$^{-1}$), with typical average velocities 
between 2000 and 3000 km.s$^{-1}$. Concerning model B1, its lower initial 
mass leads to a higher mass of the accreted envelope and thus to a higher 
ejected mass but with a lower mean velocity ($\sim$ 1000 km.s$^{-1}$), which 
translates into a lower kinetic energy. This makes this model more 
opaque to the $\gamma$-rays, as will be shown below. Concerning the 
synthesis of radioactive nuclei, the most important result is that \be7 
is mainly produced in CO novae whereas \na22 is only produced in ONe novae. 
This result is a consequence of the different initial compositions of 
CO and ONe novae, as discussed above.
The complete profiles of density, velocity and chemical composition obtained 
with the hydrodynamical code, taken when the unaccelerated expansion begins, 
have been used as inputs for the calculations performed with the Monte-Carlo 
code developed to obtain the gamma-ray emission of those classical novae. 

\begin{table*} 
\centering
\begin{minipage}{100mm}
\caption{Main properties of the ejecta one hour after peak temperature.} 
\label{Tab1}
\begin{tabular}{@{}lccccccc}
 Model & M$_{\rm wd}$    & M$_{\rm ejec}$  & $<\rm E_{\rm k}>$ & 
         \be7            & 
         \n13            & \ff18           & \na22             \\
    A1 & 1.15            & 1.8~10$^{-5}$   & 3.1~10$^{16}$     & 
         $\sim$ 0        & 
           5.5~10$^{-9}$ & 7.1~10$^{-8}$   & 9.8~10$^{-10}$    \\
    A2 & 1.25            & 1.6~10$^{-5}$   & 3.3~10$^{16}$     & 
         1.2~10$^{-11}$  & 
         2.9~10$^{-8}$   & 6.7~10$^{-8}$   & 1.6~10$^{-9}$     \\
    B1 & 0.8             & 6.3~10$^{-5}$   & 8~10$^{15}$       & 
         7.8~10$^{-11}$  & 
         1.6~10$^{-7}$   & 1.7~10$^{-7}$   & $\sim$ 0          \\
    B2 & 1.15            & 1.4~10$^{-5}$   & 3.2~10$^{16}$     & 
         1.1~10$^{-10}$  & 
         1.3~10$^{-8}$   & 3.6~10$^{-8}$   & $\sim$ 0          \\
\end{tabular}

\medskip
M$_{\rm wd}$, total ejected mass (M$_{\rm ejec}$) and ejected mass of each 
isotope are in \msun, and the average kinetic energy $<\rm E_{\rm k}>$ 
is in erg g$^{-1}$.
Models A1 and A2 correspond to ONe novae, whereas B1 and 
B2 are CO ones.
\end{minipage}
\end{table*}

\section{Gamma--ray production and propagation: Monte-Carlo Simulation}

In order to simulate the evolution of the $\gamma$-ray spectra of  
the different models of novae considered, we have developed a 
$\gamma$-ray transfer code on the basis of the method described by 
Pozdnyakov, Sobol \& Sunyaev (1983) and Ambwani \& Sutherland (1988).
This code has been  successfully applied to the 
study of the spectra of type Ia supernovae (G\'omez-Gomar, Isern \& Jean 
1997). It is based on the 
Monte-Carlo technique, which allows to treat the comptonization 
of high energy photons without approximations. In the code, the radioactive 
decays of \be7, \n13, \ff18 and \na22 are included to generate the 
initial photons.
A comment must be made regarding \be7. This nucleus decays via the capture of 
an electron of the K-shell.
This implies that at high temperatures its decay time in the laboratory could 
change because of ionization. However, as the 
lifetime of this nucleus is quite long we assume that for most of the time the 
temperature of the envelope is low enough to avoid complete ionization and 
the decay time is always taken to be equal to the laboratory time.

Once photons are generated according to the relative isotopic abundances 
and rates of disintegration of the above mentioned radioactive nuclei, 
their trip across the expanding ejecta is 
simulated by taking into account the three different interactions which 
affect their propagation, i.e.
Compton scattering, photoelectric absorption and production of 
$e^{-}$--$e^{+}$ pairs. The cross sections for all these interactions is 
computed taking into account the precise composition of the ejecta. In the 
case of Compton scattering the cross section adopted is the usual 
Klein-Nishina 
expression, while absorption and pair production cross sections are taken from 
the compilation of experimentally evaluated data maintained by the Brookhaven 
National Laboratory. In the code the effects of the expansion velocity are 
included an thus realistic line profiles are obtained. 
Besides direct emission by radioactive decay, photons are also emitted by 
positron annihilation.

The treatment of positron annihilation has required particular attention, 
since during some phases of the nova evolution this is the dominant emission 
mechanism, being \n13, \ff18 and \na22 the main contributors. When a positron 
is emitted,  either it escapes without interaction with the ejecta or  
it annihilates. We have considered that in a nova envelope positrons 
thermalize before annihilating. This approximation is wrong in less 
than $1\%$ of the cases in an electronic plasma (Leising \& Clayton 1987). 
Concerning a neutral envelope, 
the excitation cross section dominates any other interaction 
above energies of $\sim$100 eV (Bussard, Ramaty \& Drachman 1979) 
and thus positrons lose energy until they reach this extremely low value.
To reproduce this braking effect in the simulations, positrons propagate until 
they cross an equivalent column mass density of 0.2 g cm$^{-2}$ (measured in 
straight line) (Chan \& Lingenfelter 1993). This is the mean range expected 
for a 0.6 MeV positron braked down to energies $\sim$100 eV through elastic 
collisions with the surrounding medium, and neglecting 
the effects of the magnetic field on its propagation. 
When a positron is thermalized, it covers a negligible distance 
and then 
annihilates. For temperatures and densities of a nova envelope the positrons 
form positronium in $\sim$90 \% of annihilations (Leising \& Clayton 1987) 
while in the 
remaining 10 \% of cases they annihilate directly. 25\% of times positronium 
is formed in singlet state, which leads to the emission of two 511 keV photons, 
while 75\% of times it is formed in triplet state, leading to a three photon 
annihilation. 
The continuum spectrum of photons produced by triplet annihilations is 
obtained from the results of Ore \& Powell (1949).
To summarize, once a positron is produced we follow its trip until it 
escapes or covers the average energy loss distance. In the 
latter case it 
produces positronium 90\% of times leading to triplet to singlet annihilations 
in the proportion 3:1, while in 10 \% of the cases it annihilates directly. 

\section{Gamma-ray spectra and light curves}

The $\gamma$-ray signatures of CO and ONe novae are very different. In the 
CO case, an important production of \be7 ensues (Hernanz et al. 1996), 
leading to line emission at 
478 keV. There is also line emission at 511 keV, related to 
e$^-$--e$^+$ annihilation, 
and a continuum related to comptonized 511 keV emission and positronium decay 
(see below for details). On the other hand, ONe novae are \na22 producers, 
leading to line emission at 1275 keV (plus the 511 keV and below emission 
similar to the one of the CO cases). 

In figures 1, 2, 3 and 4 the spectral evolution 
of models A1, A2, B1 and B2, representing the emission expected for these 
novae at a distance of 1 kpc, is shown. These sequences correspond to 
the early stages of the 
nova outburst up to 2 days after peak temperature. In all cases a continuum 
component, basically below 511 keV, as well as some lines (478 keV, 511 keV 
and 1275 keV) appear. The relative intensity of the continuum and the lines, 
the particular lines appearing and the temporal evolution of the spectra 
depend on the model considered. In what follows we will discuss in detail  
the properties of the continuum and the lines.

\begin{figure}
	\begin{center}
\epsfig{file=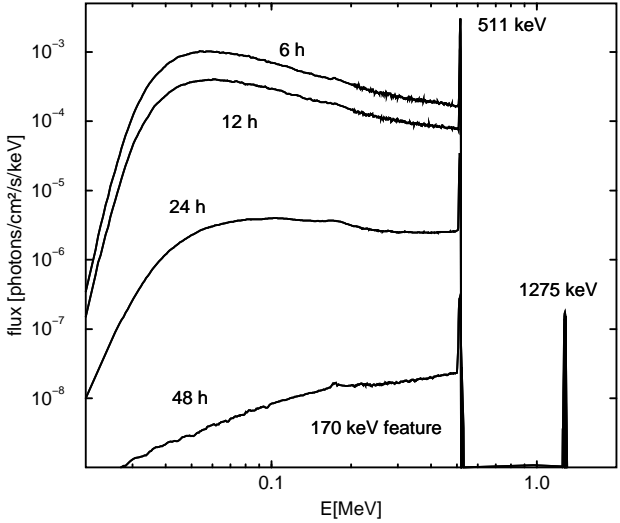,height=7 cm}
	\end{center}
\caption{Evolution of the $\gamma$-ray spectrum of model A1 
(1.15 \msun, ONe nova; d=1 kpc).
Spectra correspond to 6, 12, 24 and 48 hours after peak temperature.
Notice that the 1275 keV line has not yet reached its maximum flux at 
t=48 h (see Table 3 and Figure 6).}
\label{fig1}
\end{figure}

\begin{figure}
	\begin{center}
\epsfig{file=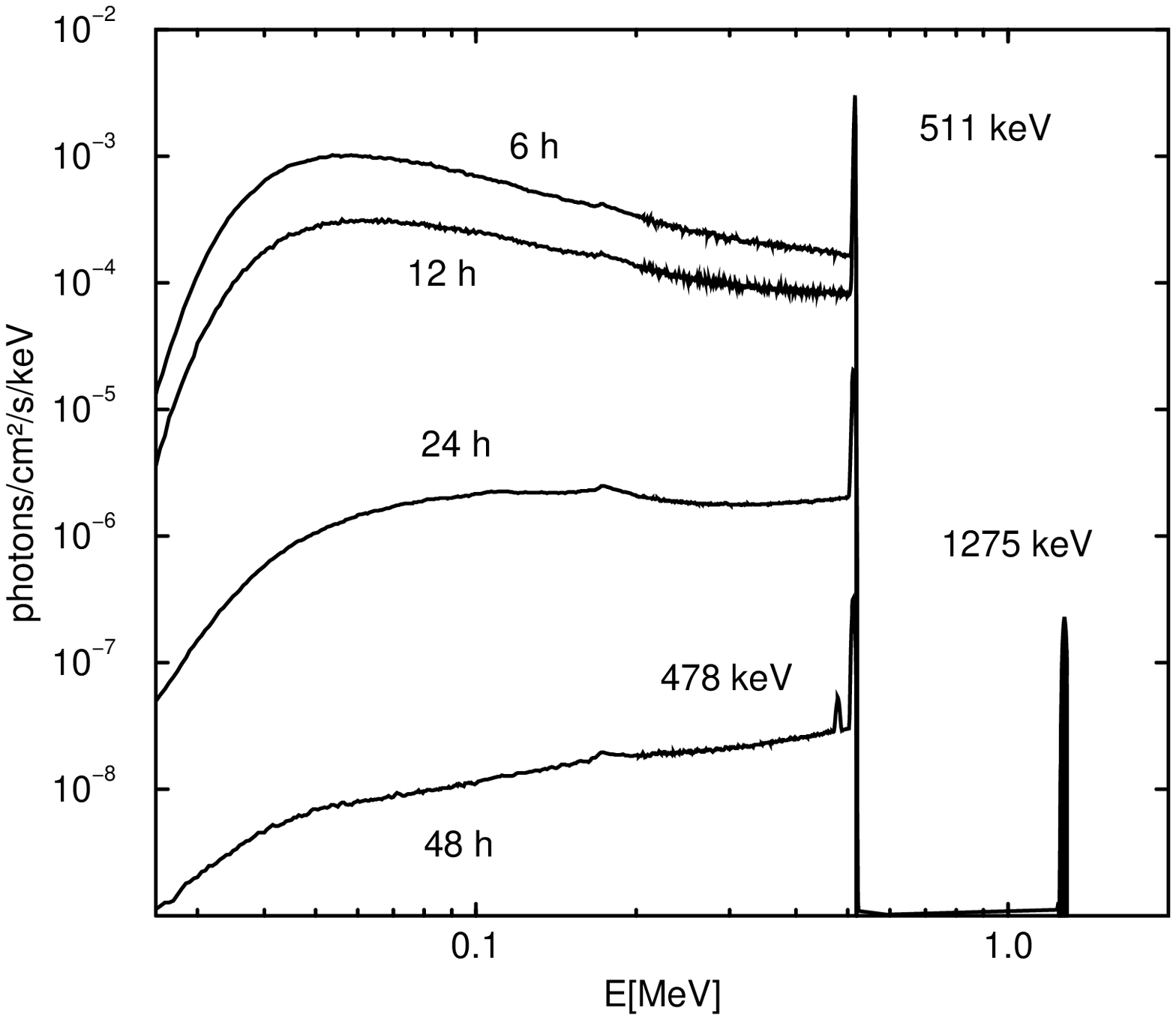,height=7 cm}
	\end{center}
\caption{Evolution of the $\gamma$-ray spectrum of model A2 
(1.25 \msun, ONe nova; d=1 kpc).
Spectra correspond to 6, 12, 24 and 48 hours after peak temperature.}
\label{fig2}
\end{figure}

\begin{figure}
\begin{center}
\epsfig{file=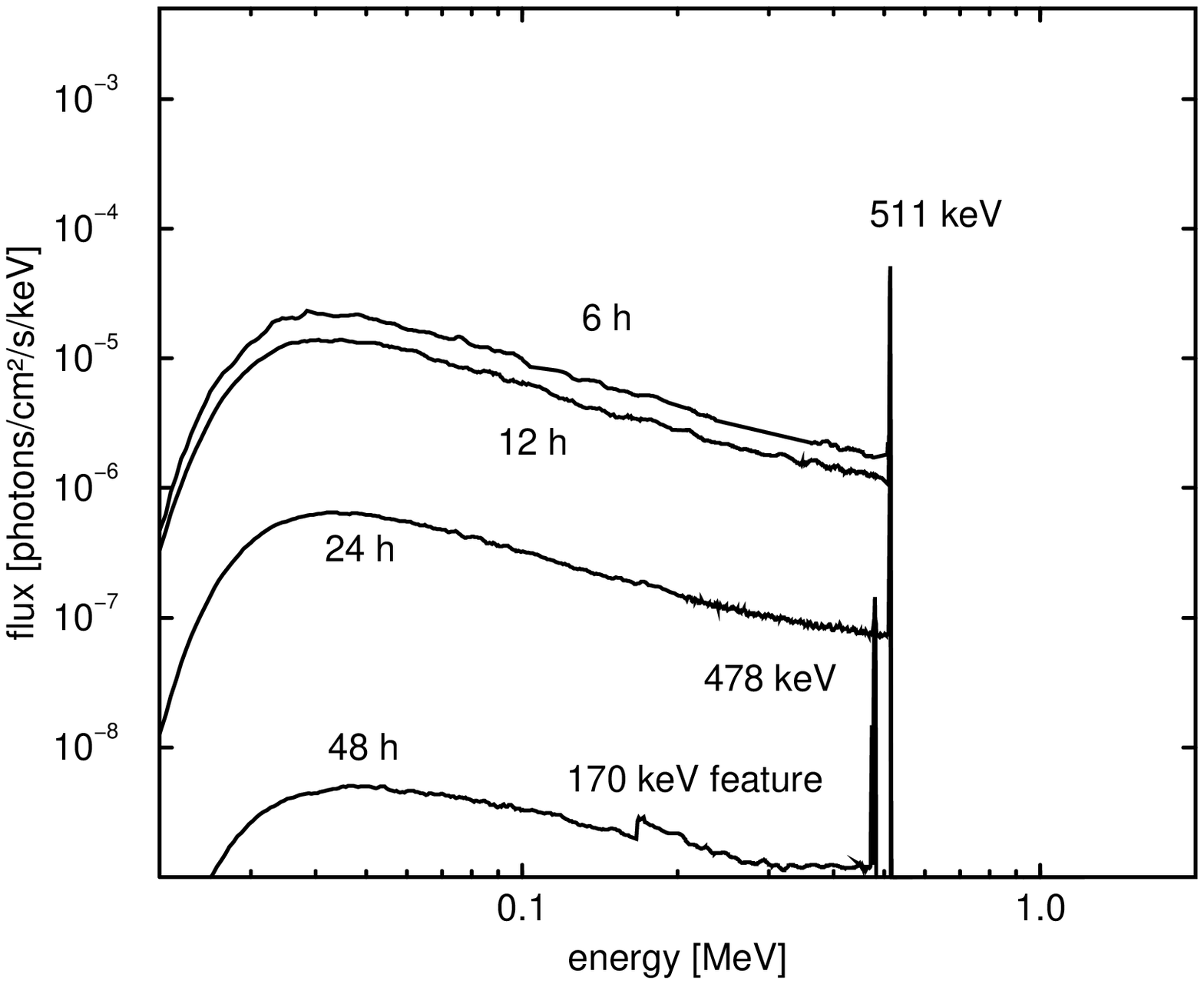,height=7 cm}
\end{center}
\caption{Evolution of the $\gamma$-ray spectrum of model B1 
(0.8 \msun, CO nova; d=1 kpc).
Spectra correspond to 6, 12, 24 and 48 hours after peak temperature.}
\label{fig3}
\end{figure}

\begin{figure} 
	\begin{center}
\epsfig{file=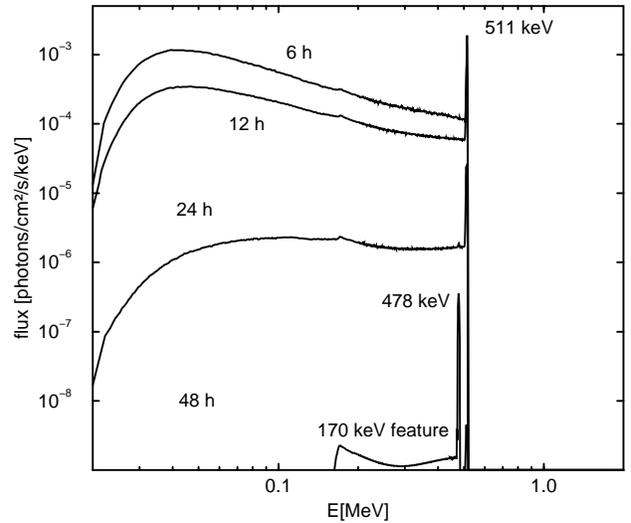,height=7 cm}
	\end{center}
\caption{Evolution of the $\gamma$-ray spectrum of model B2 
(1.15 \msun, CO nova; d=1 kpc).
Spectra correspond to 6, 12, 24 and 48 hours after peak temperature.}
\label{fig4}
\end{figure}

\subsection{Continuum}

\begin{figure}
	\begin{center}
\epsfig{file=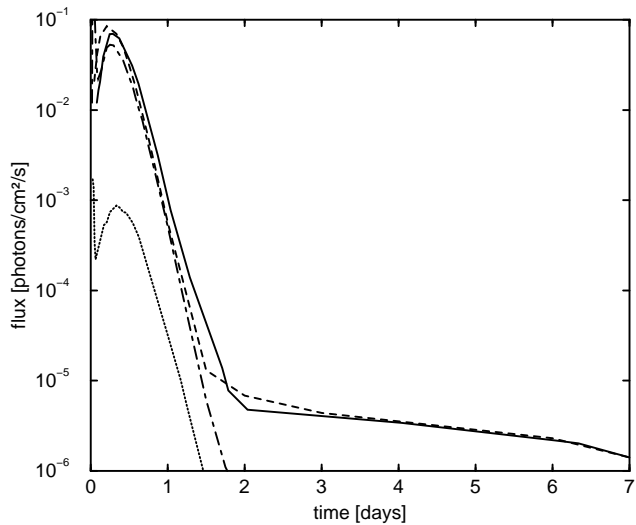,height=7 cm}
	\end{center}
\caption{Evolution of the $\gamma$-ray emission in the band 170 keV--470 keV 
(d=1 kpc) for the different models. Solid line corresponds to model A1 
(1.15 \msun, ONe nova), dashed line to model A2 (1.25 \msun, ONe),
dotted line to model B1 (0.8 \msun, CO), and dot-dashed line to 
model B2 (1.15 \msun, CO).}
\label{fig5}
\end{figure}

In all models a strong continuum dominates the spectrum during the 
early epochs of the expansion, when the ejecta is optically thick.
It is produced both by the 
comptonization of photons emitted in the 511 keV line and by the photons  
directly emitted in the continuum by positronium annihilation. 
In order to display the evolution of the continuum in all models, we have 
computed the $\gamma$-ray light curves for the emission in a band ranging 
from 170 keV to 470 keV (see fig. \ref{fig5}). This band contains most 
of the continuum and no appreciable contribution of the 478 and 511 keV lines. 
Furthermore, it avoids energies where the instrumental background for the 
SPI spectrometer on board the future INTEGRAL satellite will be 
higher \cite{Pj95}. 
The duration of this period is determined by the mass of the ejecta and 
its expansion rate, being longer for more massive envelopes.
As shown in figure \ref{fig5}, a reduction of the flux by a 
factor of 100 occurs at around day 1. In this phase,
the main source of $\gamma$-ray photons is the annihilation of positrons 
coming from the decay of \n13 and \ff18. In 
all cases the continuum displays a cutoff at low energies (see figs. 1 to 4). 
During the first 
hours, the cutoff is caused by photoelectric absorption, which has larger 
cross sections than Compton scattering at low energies and acts as a sink 
of comptonized photons. Because of the strong effect of the composition on the 
cross section of photoelectric absorption, the cutoff is located at higher 
energies for the ONe rich envelopes ($\sim$30 keV to be compared to 
$\sim$20 keV for CO models). As the ejecta expands, the optical thickness of 
the envelope decreases and comptonization becomes unable to ``pump'' photons 
to low energies, leading to the shift of the cutoff towards higher energies 
and to smoother slopes. When the ejecta becomes transparent, the contribution 
of comptonization to the continuum completely disappears and the 
continuum corresponds exclusively to the photons coming from positronium 
annihilation, which displays its intrinsic emission spectrum (Ore \& Powell  
1949).
In the transition from an optically thick to an optically thin envelope, 
a feature appears at 170 keV (the energy of a backscattered 511 keV photon) 
which is related to the fact that very few photons are scattered more than 
once, leading to a steep decrease of the continuum intensity 
at this energy (see figs. 1 to 4).
In the case of CO novae (models B1 and B2), after the decay of \n13 and 
\ff18 no 
relevant e$^+$ emitters are left and the continuum practically disappears 
(flux F$\leq 10^{-6}$ photons/s/cm$^2$) between days 1 and 2.
However, for ONe novae (models A1 and A2) \na22 is a secondary 
source of e$^+$, with a longer lifetime than \n13 and \ff18. It  
emits a e$^+$ in 90 \% of its decays. For these models, the continuum would 
in principle decline with the decay time of \na22 (3.75 years); however, when 
the ejecta is so transparent that even the positrons generated by the decay of 
\na22 can escape without interacting with matter, the continuum 
disappears. This happens at t$\sim$7.5 days for models A1 and A2 (see fig. 
\ref{fig5}) .

The maximum for the emission in the continuum coincides with 
that for the 511 keV line in all cases, as we will see below 
(fig. \ref{fig8}). It appears 
around 5--6 hours after peak temperature. This maximum is mainly 
associated with the decay of \ff18 ($\tau$=2.6 hours). However, in models 
B1 and B2 a previous and stronger maximum appears around one hour 
after peak temperature, 
caused by the emission of the short-lived \n13. This maximum is 
related to the high abundance of \n13 in the outer layers of these models.
In Table \ref{Tab2} the positions and intensities of these maxima, as well 
as the moment when the continuum intensity falls below 10$^{-6}$ 
photons/s/cm$^2$ (t$_{\rm off}$) are summarized.

\begin{table*}
\centering
\begin{minipage}{120 mm}
\caption
{Positions and intensities of the flux maxima in the continuum 
(170 keV--470 keV) for the different models. Subscript 1 corresponds to 
values (if they exist) for the maxima associated with \n13 decay, while 
subscript 2 corresponds to maxima associated with \ff18 decay. 
At t$_{\rm off}$, the flux in the continuum falls below 10$^{-6}$ 
photons/s/cm$^2$. D is the maximum detectability distance for INTEGRAL's SPI 
(see text).}
\label{Tab2}
\begin{tabular}{@{}lcccccc}
Model &
t$_1$ [hours] & Flux$_1$ [photons/s/cm$^2$] & t$_2$ [hours] &
Flux$_2$ [photons/s/cm$^2$] &
t$_{\rm off}$ [days] &
D [kpc] \\
A1 & -   & - & 5.5 & 7.4~10$^{-2}$ & 7.5 & 8.8  \\
A2 & -   & - & 5 & 9.1~10$^{-2}$ & 7.5 & 8.8 \\ 
B1 & 1   & 1.7~10$^{-3}$ & 8.4 & 8.3~10$^{-4}$ & 1.4 & 1.1\\
B2 & 1   & 1.7~10$^{-1}$ & 6.2 & 5.7~10$^{-2}$ & 1.7 & 8\\

\end{tabular}
\end{minipage}
\end{table*}

Although the continuum duration is very short as compared 
to the nominal observation times of INTEGRAL's SPI instrument, 
its large intensity 
during this short period makes it interesting to compute the detectability 
distance for the different models. This is the value appearing in the last 
column of Table \ref{Tab2}, and it has been computed taking into 
account the detailed response of INTEGRAL's SPI \cite{Pj95}. We have assumed 
an observation lasting only 10 hours, in order to obtain good sensitivities 
when the flux in the continuum is still high. All the observations start 
5 hours after peak temperature. All the computed models, except the low-mass 
CO nova (B1), could be detected in the continuum up to the Galactic center.

\subsection{Lines}

\begin{figure}
	\begin{center}
\epsfig{file=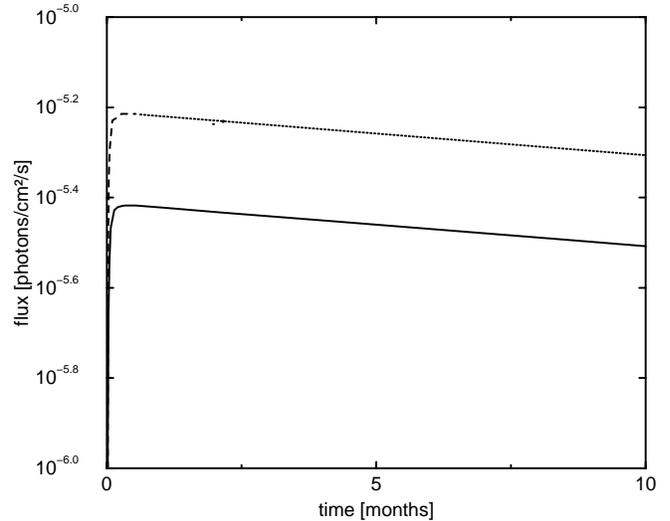,height=7 cm}
	\end{center}
\caption{Light curves for the 1275 keV line (d=1 kpc). Solid line corresponds to 
model A1 (1.15 \msun, ONe nova) and dashed line corresponds to A2 
(1.25 \msun, ONe).}
\label{fig6}
\end{figure}

Lines are the best candidates for the detection of novae. This is 
particularly true in the case of a high resolution spectrometer like 
INTEGRAL's SPI 
because a flux emitted in a narrow line is detected with these instruments
with significances much higher than if it was spread in a broad band.
The strongest  $\gamma$-ray lines emitted by CO and ONe novae are shown 
in figures 1 to 4, where the spectral 
evolution of models A1, A2, B1 and B2 during the first 2 days is displayed. 
Two phases can be distinguished in the  evolution of all lines. During 
the early nova expansion (up to a week, depending on the model), the envelope 
is optically thick and the intensity is not only determined by the 
instantaneous 
abundance of the parent isotopes but also by their distribution as a function 
of the optical depth (which depends on densities and expansion velocities).
At this stage, the evolution of lines is affected by two timescales: the 
decay times of  the radioactive isotopes and the characteristic timescale 
of envelope expansion. This phase includes both the rise of the 
intensity of the long-lived 1275 keV and 478 keV lines, and the 
rise and decay of the 511 keV line. Later on, when  the effects of absorption 
and comptonization become negligible, the intensity of the lines is 
determined by the total mass of the radioactive isotopes and the 
evolutionary timescale of the 1275 keV and the 478 keV lines simply 
corresponds to the decay times of \na22 and \be7 (see figures 
\ref{fig6} and \ref{fig7}, where 
light curves for the 1275 and 478 keV lines are shown for ONe and CO novae, 
respectively).
However, even at late times line profiles (which have to be taken into 
account to determine the detectability of the lines)
are determined by the distribution of \na22 and \be7. 

\subsubsection{The 478 keV and 1275 keV lines}
The observation of any of these lines would allow to discriminate  
between ONe (A1 and A2) and CO novae (B1 and B2). While the 
1275 keV line can only be observed in the case of ONe novae (a negligible 
\na22 amount is produced by CO novae), the reverse is true for the 478 keV 
line, except for model A2, where a small amount of \be7 (see Table 1) leads 
to a weak line at 478 keV (see figures 1 to 4).
In Table \ref{Tab3}, 
we summarize the epoch of the maximum, the total line flux at maximum and the 
detection distances for the lines at 3 $\sigma$ significance. These detection 
distances have been computed for INTEGRAL's SPI assuming an observation time 
of 10$^6$s starting at maximum. In all cases, the detection distances are 
$\sim$0.5 kpc.
For the computation of the detection distances, we have 
taken into account the profiles of the lines: typical widths 
(FWHM) when the envelope is transparent are in the range 
between 3 keV (model B1) and 7 keV (model B2), for the 478 keV 
line, and around 20 keV for the 1275 keV one (for both A1 and A2 
models).

The rise phase for the 1275 keV line lasts $\sim$7 days. 
Soon after the maximum, the light curve reaches the stable 
decline phase with the expected decay time of 3.75 years. During this phase, 
the line  intensities are directly determined from the ejected \na22 mass 
of each model, which grows with the mass of the progenitor white dwarf 
(Table \ref{Tab1}), a tendency that had been previously pointed out by other 
authors (Starrfield et al. 1993, Politano et al. 1995). 

\begin{figure}
	\begin{center}
\epsfig{file=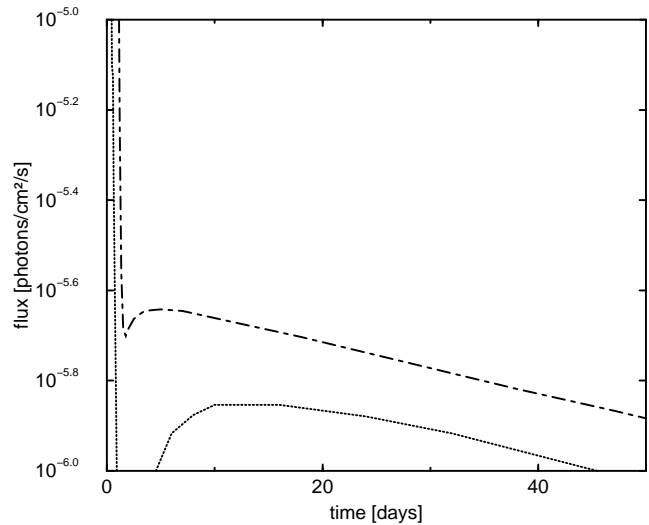,height=7 cm}
	\end{center}
\caption{Light curves for the 478 keV line (d=1 kpc). Dotted line 
corresponds to model B1 (0.8 \msun, CO nova), whereas 
dot-dashed line corresponds to model B2 (1.15 \msun, CO).} 
\label{fig7}
\end{figure}

On the basis of the calculations by Starrfield et al. (1978), Clayton (1981) 
first suggested that the 478 keV line produced by the decay of \be7 could 
be detectable for a nearby nova. This possibility was subsequently discarded 
by the results obtained by Boffin et al. (1993), who considered an updated 
reaction network, but did one-zone model nucleosynthesis calculations. 
The new calculations performed by Hernanz et al. (1996) show that 
convection plays an important role and that an appreciable amount 
of \be7 is produced by a CO nova. The possible detection of the 
corresponding 478 keV line would be an important confirmation of the 
thermonuclear runaway model for novae and its associated nucleosynthesis. 
In fig. \ref{fig7} we distinguish two phases in the evolution of 
the 478 keV line for the two CO models. During the first $\sim$1.5 days after 
peak temperature, the 
intensity of the line is completely dominated by the contribution of the 
very strong continuum generated by \n13 and \ff18 emission. 
During this epoch, the line reaches its absolute maximum.
Later on, the contribution of the continuum disappears 
and the line follows a ``typical'' light curve. The properties of the line 
at its late maximum are shown in Table \ref{Tab3}. This 
late maximum is reached at day 5 for the massive CO nova B2 
(F=2.6$~$10$^{-6}$ photons/s/cm$^2$). The low-mass CO nova, B1, has a less 
intense maximum (F=1.4$~$10$^{-6}$ photons/s/cm$^2$), related to its lower 
\be7 content, appearing later (t$\sim$13 days), because it is more opaque.

\begin{figure}
	\begin{center}
\epsfig{file=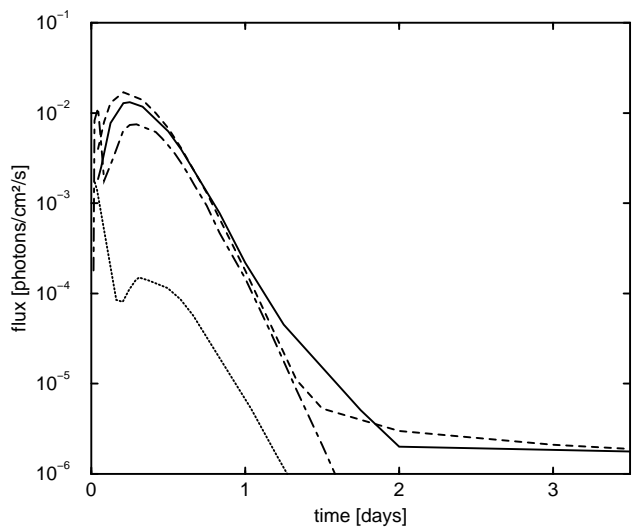,height=7 cm}
	\end{center}
\caption{Light curve evolution for the 511 keV line (d=1 kpc). 
Solid line corresponds to model A1 (1.15 \msun, ONe nova),
dashed line to model A2 (1.25 \msun, ONe), dotted line to model 
B1 (0.8 \msun, CO) and dot-dashed line to model 
B2 (1.15 \msun, CO).}
\label{fig8}
\end{figure}

\begin{table*}
\centering
\caption
{Properties of the 478 keV and 1275 keV lines: time and flux 
at maximum for an assumed distance of 1 kpc. D is the maximum 
detectability distance for INTEGRAL's SPI (see text).}
\label{Tab3}
\begin{tabular}{@{}lcccc}
Model & line     & t [days] & Flux [photons/s/cm$^{2}$] & D [kpc]  \\
A1    & 1275 keV & 7.5      & 3.8~10$^{-6}$             & 0.5 \\
A2    & 1275 keV & 6.5      & 6.1~10$^{-6}$             & 0.6 \\ 
B1    &  478 keV & 13       & 1.4~10$^{-6}$             & 0.4 \\
B2    &  478 keV & 5        & 2.6~10$^{-6}$             & 0.5 \\
\end{tabular}
\end{table*}

\subsubsection{The 511 keV line}

\begin{table*}
\centering
\caption
{Properties of 511 keV line: time and flux 
at maximum for an assumed distance of 1 kpc. D is the maximum 
detectability distance for INTEGRAL's SPI (see text).}
\label{Tab4}
\begin{tabular}{@{}lccc}
Model  & t [hours] & Flux [photons/s/cm$^{2}$]   & D [kpc]  \\
A1     & 6         & 1.3~10$^{-2}$               & 13.5 \\
A2     & 5         & 1.6~10$^{-2}$               & 12.5 \\ 
B1     & 1/7.5     & 1.7~10$^{-3}$/1.4~10$^{-4}$ & 1.6 \\
B2     & 1/6.5     & 10$^{-2}$/7.3~10$^{-3}$     & 9 \\
\end{tabular}
\end{table*}

This line is expected to be, by far, the strongest $\gamma$-ray feature 
emitted by any nova and it appears with high intensity in all models. 
However, it is much fainter in the low-mass CO model (B1), which has a 
more opaque envelope because of its higher mass and lower velocity. 
Several isotopes contribute to the 511 keV emission (see section 1), 
with decay times ranging from minutes to years. This results in a relatively 
complex temporal evolution. Besides, as the line reaches its maximum during 
the 10 first hours after peak temperature, when the envelope is optically 
thick, its intensity is strongly affected by the details of the distribution 
of the radioactive nuclei and the expansion velocities. Thus its detection 
would provide critical information about the properties of novae 
ejecta. The properties of the line in the different models appear  
in Table \ref{Tab4}, whereas the light curves are displayed in figure 
\ref{fig8}.
As happens with the continuum, the 511 keV line shows two maxima in models 
B1 and B2, which are related 
to the decay of \n13 and \ff18, while the first maximum does not appear in 
models A1 and A2. The maximum intensities are shown in Table \ref{Tab4}. 
Light curves for models 
A1 and A2 are very similar, specially in the period between 0.5 days and 
1 day after peak temperature. The intensity of the late maximum for model B2 
is approximately 2 times smaller than that of models A1 and A2, 
although later than 12 hours after peak temperature its light curve converges 
with those of models A1 and A2. Model B1 emits lower fluxes at the two 
maxima (around 100 times smaller than ONe models for the 
second maximum), because of the 
larger optical thickness at these early epochs. 
Later than 2 days after peak temperature, models A1 and A2 present a more 
stable emission at 511 keV produced by the contribution of \na22 
which emits a e$^+$ in  90 \% of its decays. 
As happens for the continuum (see above), this remaining emission totally 
disappears later than a week after peak temperature, when 
positrons emitted by \na22 freely escape without annihilating. 

For the estimation of  the maximum distances for detectability (see 
Table \ref{Tab4}), we have 
adopted in this case the same observation parameters as for the continuum, 
because of the shortness of the 511 keV emission (for comparison, 
the sensitivity offered by OSSE, on board CGRO satellite, at 511 keV is 
approximately 10 times worse than that offered by INTEGRAL's SPI for the same 
conditions). For models A1, A2 and B2, the line would be
observed at distances equivalent to the Galactic center (13.5 kpc, 
12.5 kpc and 9 kpc) although the Galactic center itself should be avoided 
because of its 511 keV characteristic emission. 
The lower intensity of the 511 keV line
for the low-mass CO nova, B1, makes it only detectable for a distance 
smaller than 1.6 kpc.
The typical width (FWHM) of the 511 keV line is 8 keV (except 
for model B1, which has a width of 3 keV).

\begin{figure}
	\begin{center}
\epsfig{file=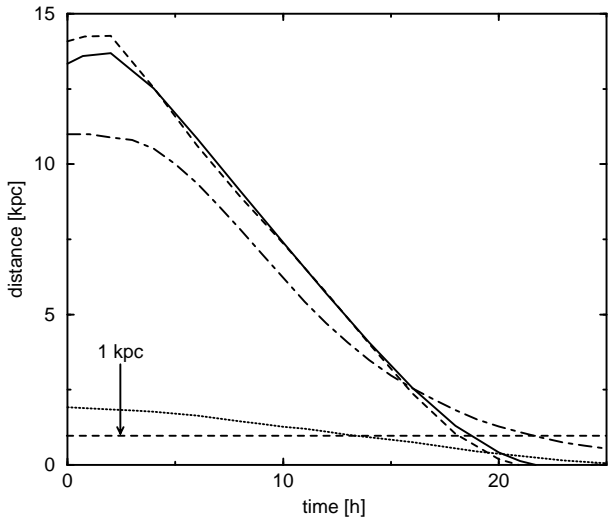,height=7 cm}
	\end{center}
\caption{Maximum distance of detection for the 511 keV line emission as a 
function of the starting observation time. Solid line corresponds to 
model A1 (1.15 \msun, ONe nova),
dashed line to model A2 (1.25 \msun, ONe), dotted line to model 
B1 (0.8 \msun, CO) and dot-dashed line to 
model B2 (1.15 \msun, CO). The results correspond to a 10 hours 
observation with INTEGRAL's SPI.}
\label{fig9}
\end{figure}

The fundamental difficulty posed by the detection of the 511 keV line is, 
of course, its short duration. The  detectability of the line critically 
depends on the time at which the observation starts. To visualize this 
dependence, we have computed the maximum distance of detection for a 
10 hours observation as a function of the starting time of the observation 
(Fig. \ref{fig9}). It is easy to appreciate from the figure how fast the 
chances of detection decline. The maximum distance of detection falls 
below 1 kpc if the observation starts later than $\sim$20 hours 
after peak temperature.  
Thus a strategy for early detection of novae is necessary in order  
to observe this emission. A chance is offered by the fact that 
INTEGRAL will spend an important part of its observation time performing 
a survey of the Galactic plane, where most of the novae are expected to 
occur. Because of its large field of view ($\sim16^{\rm o}$), the 
possibility of an early nova detection by INTEGRAL's SPI during the survey 
is not negligible.

\section{Discussion and Conclusions}
It is important to mention that all theoretical models, including ours, do 
not fit some of the ejected masses observed in novae eruptions (although 
not all observations of the same nova give the same value of the ejected 
mass). As an example, our ONe model A1 (M$_{\rm wd}$=1.15) fits quite well 
the observed abundances of the neon nova QU Vul 1984, but the ejected mass 
of this nova deduced from observations is between 2$\times 10^{-5}$ and 
$\sim10^{-3}$ \msun (Saizar et al. 1992), almost two orders of magnitude 
larger than the theoretical one in the most unfavourable case. 
Thus, the flux of the 1275 keV line could be considerable larger for some 
particular ONe nova, but no present theoretical models are able to produce 
simultaneously such large ejected masses and neon in the ejecta (see 
Shara 1994, Shara \& Prialnik 1994 and Starrfield et al. 1997 for recent 
discussions and some suggestions relative to this problem). In view of 
this still open problem, we have made some estimations of the effect of the 
mass of the envelope, as well as the mean velocity of the ejecta, in the flux 
of the predicted $\gamma$-ray lines. Concerning the 511 keV line, an increase 
by a factor of $\sim$20 in the ejected mass only reduces the flux at maximum 
by a factor of $\sim$3. On the other hand, a reduction of the speed by a 
factor of $\sim$8 reduces the flux by a factor of $\sim$100. Concerning the 
medium and long lived-emission from \be7 and \na22 (at 478 and 1275 keV, 
respectively), ejected mass influences in the expected way the fluxes (a 
factor of 10 increase in mass leads to a factor 10 increase in flux), because 
that emission is mainly produced when the envelope is transparent. By the same 
reason, the velocity only influences the rise to maximum, because this is the 
phase where transparency plays an important role.

One of the main conclusions of our study is that the $\gamma$-ray signatures 
of CO and ONe novae are very different, except for the early emission. During 
the first hours after explosion (i.e. after peak temperature or roughly after 
peak bolometric luminosity), continuum emission below 511 keV dominates, as  
well as an intense line at 511 keV, for all the models. But this emission can 
more 
easily be detected ``a posteriori'', with a retrospective analysis of the 
data from instruments like BATSE on board CGRO (Fishman et al. 1991). 
The reason is that this early emission 
happens most probably before the maximum in visual magnitude and even before 
detection for the majority of novae: around 2 days before visual maximum for 
a very fast nova, between 2 and 5 days for a fast nova and between 5 and 10 
days before maximum for a moderately fast nova; the majority of ONe novae 
have been classified as fast novae. We are mainly concerned by moderately fast 
to very fast novae, because these are the most luminous ones and thus also the 
most easily detectable either in visual or in $\gamma$-rays. Concerning the 
late emission, it is very different in a CO nova than in an ONe one. In the 
first case, emission from \be7 at 478 keV is obtained, with fluxes lasting 
for some months but just in the limit of detectability by present and future 
instruments (see however our comment concerning ejected masses, which applies 
both to CO and to ONe novae). A similar situation concerning detectability 
occurs in the ONe case, but related to the 1275 keV line, which lasts around 
3 years. We understand why neon novae have not been detected by COMPTEL, and 
we are moderately optimistic concerning the future SPI instrument on board 
INTEGRAL, specially if a nearby and high envelope mass nova explodes during 
the duration of the mission.

\section*{Acknowledgments}
We are specially indebted to Pierre Jean, who provided us with information 
about the SPI spectrometer of the future INTEGRAL satellite.
We thank for partial support the CICYT and DGICYT Projects ESP95-0091 and 
PB94-0827-C02-02.

\label{lastpage}

\end{document}